%Paper: cond-mat/9510004
%From: Okunishi <okunishi@godzilla.phys.sci.osaka-u.ac.jp>
%Date: Mon, 2 Oct 95 12:27:37 +0900

\magnification = \magstep1
\baselineskip = 22pt plus 2pt minus 1pt
\hsize 14truecm

\font\fontA=cmss10 scaled\magstep3
\font\fontB=cmr10
\font\fontC=cmssi10

\fontB

%1234567890123456789012345678901234567890123456789012345678

\centerline{\fontA Product Wave Function }
\centerline{\fontA Renormalization Group }
\vskip 15pt
\centerline{ T. Nishino$^{1)}$  and K. Okunishi$^{2)}$ }
\centerline{\it $(1)$ Physics Department, Graduate School
of Science, }
\centerline{\it Tohoku University, Sendai 980-77}
\centerline{\it $(2)$ Physics Department, Graduate School
of Science, }
\centerline{\it Osaka University,  Toyonaka, Osaka 560}
\vskip 20pt
\centerline{(~~~~~~~~~~~~~~~~~~~~~~~~~~~~~~~~~~~~~~~~~)}
\vskip 10pt
\centerline{\fontC Synopsis}
\vskip  5pt

{\narrower
The authors propose a fast numerical renormalization
group method --- the product wave function renormalization
group (PWFRG) method --- for 1D  quantum lattice models
and 2D classical ones.  A variational wave function, which
is expressed by a matrix product, is improved through a
self-consistent calculation.  The new method has the same
fixed point as the density matrix  renormalization group
method. }

\vskip 1truecm
\centerline{--- to appear in J. Phys. Soc. Jpn. ---}

\vfill\eject

The real-space renormalization group$^{1)}$ (RSRG) is a
basic concept in statistical physics. One of the recent
progress in the RSRG is the density matrix renormalization
group (DMRG) method established by White.$^{2,3)}$ The
DMRG method has been applied to various one-dimensional
(1D) quantum lattice models, such as the spin
chains$^{3)}$ and electron systems,$^{4,5)}$  because the
method gives precise results for large scale systems. The
method is also applied to 2D  classical models.$^{6)}$
Recently, fixed point structure of the DMRG method has
been analyzed by \"Ostlund and Rommer.$^{7)}$ They show
that the ground state wave function obtained by the DMRG
method is a product of matrices.

Quite recently, the authors have shown that the DMRG
method is essentially the same as Baxter's variational
method on the corner transfer matrix.$^{8)}$ On the basis
of this fact, we have formulated a fast numerical method
--- the corner transfer matrix renormalization group
(CTMRG) method$^{9)}$  --- for 2D classical systems. The
CMTRG method is applicable to 1D quantum lattice models
with the help of the Trotter formula.$^{10)}$

In this paper, we propose a new renormalization group
method, which is an extension of the CTMRG method. The new
method is applicable to 1D quantum systems with no
reference of the Trotter formula.$^{10)}$ Since the method
renormalizes a wave function, which is expressed by a
matrix  product,$^{11)}$ we call it `product wave function
renormalization group (PWFRG) method.' Since the PWFRG
method and the DMRG method have many aspects in common, we
review the DMRG method at first. We then show the
numerical algorithm of the PWFRG method for 2D classical
models. We accelerate the numerical calculation with the
help of the modified Lanczos method.$^{12,13)}$ Finally we
discuss the way how to apply the PWFRG method to 1D
quantum systems.

We choose the `interaction round a face (IRF) model' as
an example of the 2D classical model.$^{8)}$ The IRF model
is defined by a Boltzmann weight $W(a'b'|ab)$ on each
face, that is surrounded by four $n$-state spins a, b, a',
and b'. The row-to-row transfer matrix is
$$
T(a'b'c' \ldots y'z'|abc \ldots yz)
= W(a'b'|ab) W(b'c'|bc) \ldots W(y'z'|yz) ,
\eqno(1)
$$
where the variables $\{a'b'c' \ldots y'z'\}$ and $\{abc
\ldots yz\}$ denote the $n$-state spins in subsequent
rows. We assume that $W(ab|cd)$ is symmetric --- $W(ab|cd)
= W(ba|dc)$ $= W(ca|db) = W(dc|ba)$ --- in order to
simplify the discussion. Generalizations for asymmetric
cases are straightforward.

The DMRG method maps the transfer matrix $T$ in Eq.1
into the effective one$^{6)}$
$$
{\tilde T}(\xi'i'j'\eta'|\xi ij\eta)
= P(i'\xi'|i\xi) W(i'j'|ij) P(j'\eta'|j\eta) ,
\eqno(2)
$$
where $P$ represents the effective transfer matrix for
the left/right-half lattice. The greek indices
$\xi,\xi',\eta$ and $\eta'$ denote $m$-state block-spin
variables, that are shown by squares in Fig.1. The
eigenvector $V$ that corresponds to the largest eigenvalue
of ${\tilde T}$ is well approximated by a matrix
product$^{2,3,7)}$ (Fig.1)
$$
V(i\xi|j\eta) = \sum_{\alpha}
R(i\xi|\alpha) \, \omega_{\alpha}^{~} \,
{R_{~}^T}(\alpha|j\eta) ,
\eqno(3)
$$
where $R$ is a $nm$ by $m$ orthogonal matrix, $R^T$ is
its transpose, and the relation
$
\sum_{i\xi} {R^T}(\alpha|i\xi) R(i\xi|\beta)
= \delta_{\alpha}^{\beta}
$
is satisfied. We have expressed the $(nm)^2$-dimensional
vector $V$, which we call `product wave function,' as a
$nm$-dimensional square matrix. The r.h.s. of Eq.3 is
actually the singular-value decomposition of the l.h.s of
Eq.3,  and therefore $\omega_{\alpha}$ is the singular
value (or the eigenvalue) of the `matrix' $V(i\xi|j\eta)$.

The PWFRG method gives the fixed point values of $P$ and
$R$ --- those in the thermodynamic limit --- through
successive improvements. We start from the  4-site
system.$^{2,3)}$ The initial conditions for the system
with open boundary conditions are $P = W$ (that means
${\tilde T} = W W W$), $R(i\alpha|\beta) =
\delta_{\alpha}^i \delta_{\beta}^i$,  $\omega_{\alpha} =
1/\sqrt{n}$, and $m = n$. We have to set appropriate
initial values for $P$ and $\omega_{\alpha}$ when fixed
boundary conditions are imposed. Starting from the initial
status, we improve $P$ and $R$ by way of the following
self-consistent calculations.

As a first step, we create a trial product wave function
by using $R$ and $\omega_{\alpha}$: $V(i\xi|j\eta) =
\sum_{\alpha} R(i\xi|\alpha) \, \omega_{\alpha} \,
{R_{~}^T}(\alpha|j\eta)$. The vector $V$ is not usually
the eigenvector of ${\tilde T} = P W P$ in Eq.2. We then
multiply ${\tilde T}$ on $V$ in order to obtain the
improved Ritz vector
$$
V'(i'\xi'|j'\eta')
= \sum_{\xi i j \eta} P(i'\xi'|i\xi) W(i'j'|ij)
P(j'\eta'|j\eta)  V(i\xi|j\eta)
\eqno(4)
$$
of the effective transfer matrix ${\tilde T}$, where the
inequality $(V',{\tilde T}V') / (V',V') \ge (V,
{\tilde T}V) / (V,V)$ is satisfied.

Second, we decompose $V'$ into a matrix product
$$
V'(i'\xi'|j'\eta') \rightarrow \sum_{\alpha}
A(i'\xi'|\alpha) \,
\omega_{\alpha} \, {A_{~}^T}(\alpha|j'\eta')
\eqno(5)
$$
via the singular-value decomposition (or the matrix
diagonalization), where $A$ is a $nm$ by $m'$ orthogonal
matrix. We decide the new dimension $m'$ for the
block-spin variable $\alpha \, (\le m')$ so that the new
singular values $\omega_{\alpha}$ in Eq.5 are greater than
a certain threshold.

The  third step is the renormalization of the effective
transfer matrix
$$
\sum_{jj'\eta\eta'} A^{T}(\xi'|j'\eta')
W(i'j'|ij) P(j'\eta'|j\eta) A(j\eta|\xi)
\rightarrow P(i'\xi'|i\xi) ,
\eqno(6)
$$
and that of the orthogonal matrix $R$
$$
\sum_{j\zeta\rho} A^{T}(\xi|j\zeta)
R(j\zeta|\rho) A(i\rho|\eta)
\rightarrow R(i\xi|\eta) ,
\eqno(7)
$$
where the graphical representation of these equations are
shown in Fig.2. Equation 7 is actually the renormalization
of the product wave function, because the new $R$ gives a
new product wave function. After the renormalization by
Eq.6 and Eq.7, $P$ is a $nm'$-dimensional matrix,  and $R$
is a $nm'$ by $m'$ orthogonal matrix. At this point, we
return to the first step by setting $m = m'$.

We repeat the iteration shown above until $R$ becomes
invariant under the renormalization in Eq.7; this means $A
= R$. At the fixed  point of the DMRG method, the
orthogonal matrix $R$ satisfies the relation
$
R(i\alpha|\beta) \, \omega_{\beta}^{~} =
\omega_{\alpha}^{~} \, {R_{~}^T}(\alpha|i\beta) ,
$
where the matrix $R(i\alpha|\beta) \, \omega_{\beta}^{~}$
is proportional to the square of the corner transfer
matrix.$^{9)}$ After $R$ is converged to its fixed point
value, we calculate physical quantities by using the
fixed-point value of the product wave function
$V$.$^{2,3)}$ The matrix $R$  gives correlation functions;
\"Ostlund and Rommer$^{7)}$ have shown that the largest
eigenvalue of the matrix
$
t(\alpha\beta|\gamma\eta) \equiv
\sum_i R(i\alpha|\gamma) R(i\beta|\eta)
$
is equal to unity, and the second one gives the
correlation length.

The convergence of $R$ is rather slow when the
correlation length is very long. In such a case, we modify
Eq.4 to accelerate the convergence.  In addition to $V' =
{\tilde T}V$ in Eq.4, let us create another vector $V'' =
{\tilde T}V'$.  It is apparent that a certain
linear-combination between $V'$ and $V''$ gives a better
Ritz vector. Such an improvement is called  `modified
Lanczos method.'$^{12,13)}$ What we have to do is to
substitute the best linear combination $V''' = aV' + bV''$
into the l.h.s. of Eq.5, where $a$ and $b$ are adjusted so
that  $(V''',{\tilde T} V''') / (V''',V''')$ takes its
maximum value.

The PWFRG method and the DMRG method gives the same
result at their common fixed point. The main difference
between them is that the PWFRG method renormalizes the
product wave function, while the DMRG method does not.
With the use of the wave function renormalization (Eq.7),
the PWFRG method avoid the numerical  diagonalization of
${\tilde T}$. As a result, the PWFRG method runs much
faster than the DMRG method.

As a test case, we apply the PWFRG method to the square
lattice Ising model. Figure 3 shows the nearest-neighbor
spin correlation function of the Ising  model. Numerical
error in the data when $m = 40$ is less than $10^{-7}$
outside the region $2.2 \le T/J \le 2.3$, where $J$ is the
coupling constant. The result agrees with that obtained by
the DMRG$^{6)}$ and the CTMRG$^{9)}$ methods.

Finally, we discuss the way how to apply the PWFRG method
to 1D quantum lattice models. The transfer matrix of a 2D
classical model is a product of Boltzmann factors, while
the Hamiltonian of a 1D quantum lattice model is a sum of
local operators. Thus we can apply the PWFRG method to 1D
quantum models by replacing the transfer matrix
renormalization in Eq.6 by the renormalization of the
effective Hamiltonian. Such a renormalization algorithm
for the effective Hamiltonian is already given by the DMRG
method$^{2,3)}$.

The authors would like to express their sincere thanks to
Y.~Akutsu and M.~Kikuchi for valuable discussions. T.~N.
thank to S.~R.~White for helpful  comments and discussions
about the DMRG method. The present work is partially
supported by a Grant-in-Aid  from Ministry of  Education,
Science and Culture of Japan. Most of the trial numerical
calculations were done by NEC SX-3/14R in computer center
of Osaka university.

\vfill\eject

\beginsection{}References

\item{1)} T.~W.~Burkhardt and J.~M.~J.~van Leeuwen:
{\it Real-Space Renormalization,} Topics in Current
Physics vol.{\bf 30},
(Springer, 1982), and the references there in.
\item{2)} S.~R.~White: Phys. Rev. Lett. {\bf 69} (1992)
2863.
\item{3)} S.~R.~White: Phys. Rev. {\bf B48} (1993) 10345.
\item{4)} C.~C.~Yu and S.~R.~White: Phys. Rev. Lett.
{\bf 71} (1993) 3866.
\item{5)} R.~M.~Noack, S.~R.~White, and D.~J.~Scalapino:
Phys. Rev. Lett {\bf 73} (1994) 882.
\item{6)} T.~Nishino: J. Phys. Soc. Jpn. {\bf 64} (1995)
3598; cond-mat/9508111.
\item{7)} S.~\"Ostlund and S.~Rommer: cond-mat/9503107.
\item{8)} R.~J.~Baxter: {\it Exactly Solved Models in
Statistical Mechanics,}
(Academic Press, London, 1980) p.363.
\item{9)} T.~Nishino and K.~Okunishi: cond-mat/9507087.
\item{10)} H.~F.~Trotter: Proc. Am. Math. Soc. {\bf 10}
(1959) 545.
\item{11)} R.~J.~Baxter: J. math. phys. {\bf 9} (1968)
650.
\item{12)} E.~R.~Davidson, J. Comput. Phys. {\bf 17}
(1975) 87.
\item{13)} T.~Z.~Kalamboukis: J.~Phys. {\bf A13} (1980)
57.

\vfill\eject

\beginsection{~} Figure Captions

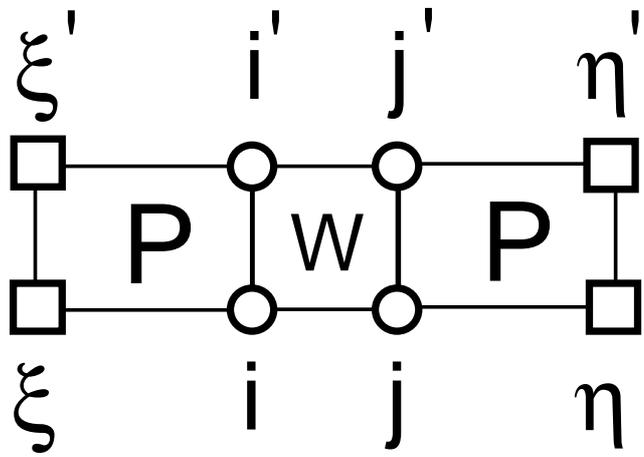
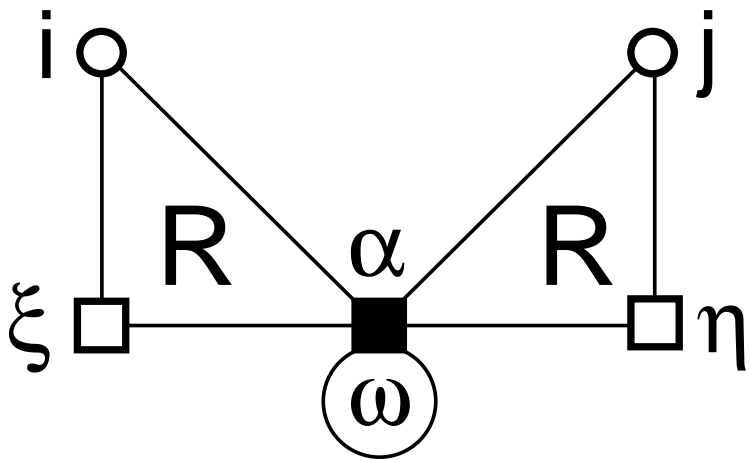
\item{Fig.~1.} Graphical representation of the effective
transfer matrix  ${\tilde T}$ in Eq.2 and the product wave
function $V$ in Eq.3.

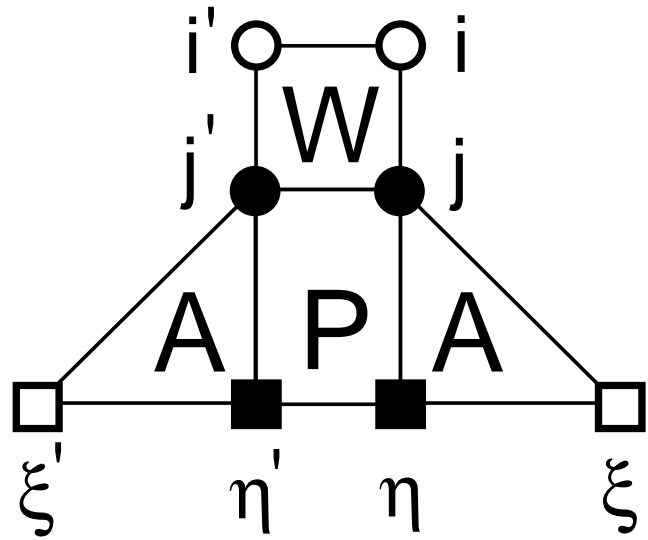
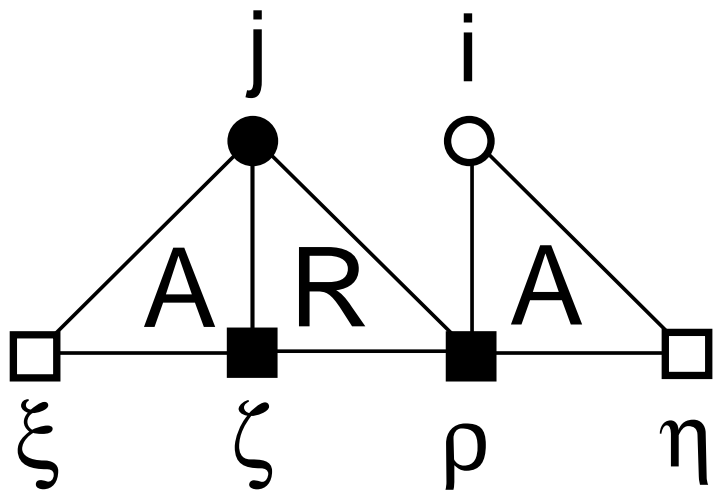
\item{Fig.~2.} Graphical representation of the
renormalization for $P$ and $R$ in Eq.6 and 7,
respectively.

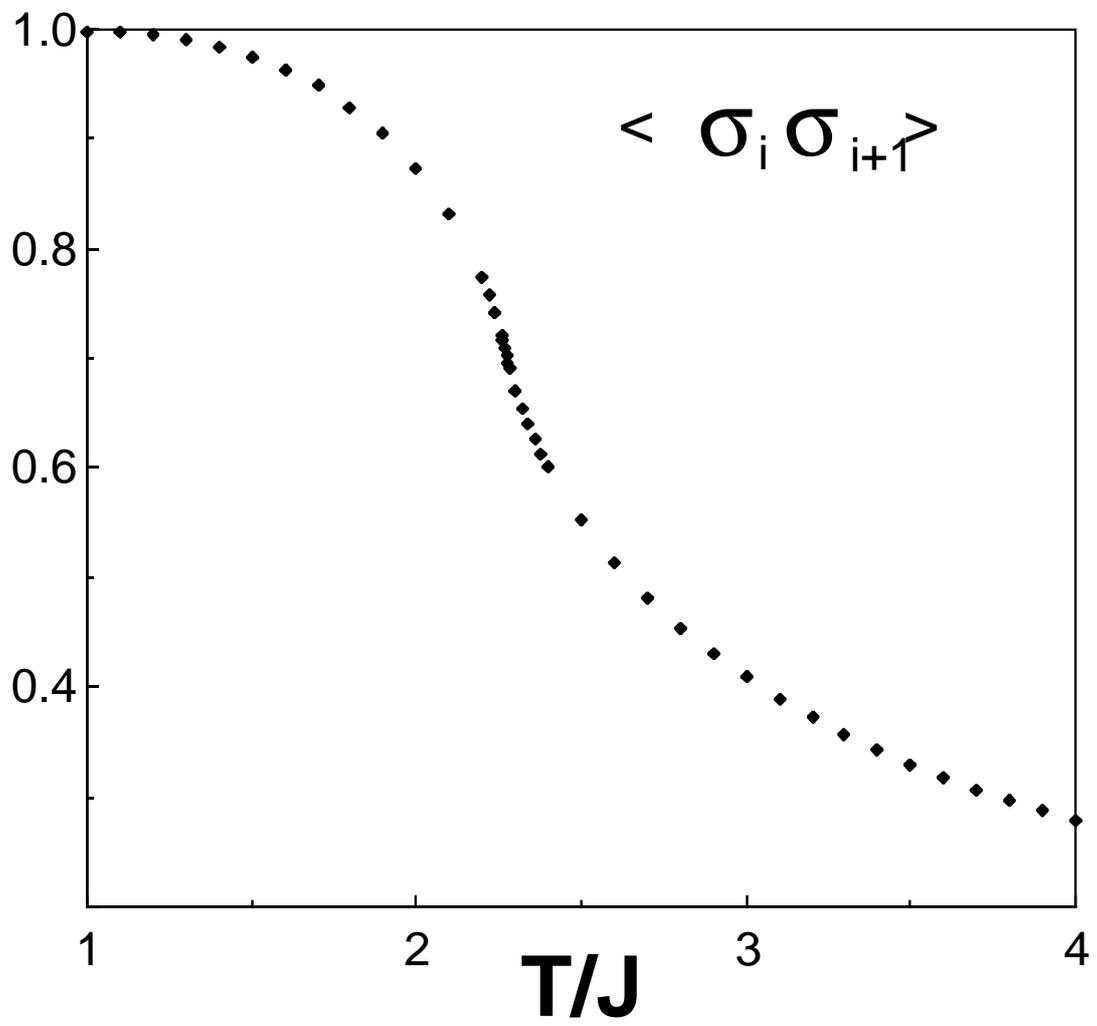
\item{Fig.~3.} Nearest neighbor spin correlation function
of the Ising model.

\vfill\bye